\documentclass{llncs}

\usepackage{etex}
\usepackage{cite}
\usepackage{threeparttablex}
\usepackage{hyperref}
\usepackage{amsfonts}
\usepackage{amssymb}
\usepackage{algpseudocode}
\usepackage{fnpct}

\usepackage{array}
\usepackage{mdwmath}
\usepackage{mdwtab}
\usepackage{eqparbox}

\usepackage[caption=false,font=footnotesize]{subfig}

\usepackage{booktabs}
\usepackage{multirow}
\usepackage{lscape}

\usepackage{listings}
\usepackage{mathtools}
\usepackage{comment}
\usepackage{xspace}
\usepackage{xcolor}
\usepackage{ulem}
\usepackage{paralist}   

\usepackage{tikz}
\usepackage{pgfplots,tikz-qtree,tikz-qtree-compat}
\usetikzlibrary{positioning}
\usetikzlibrary{pgfplots.statistics,arrows,backgrounds}
\pgfplotsset{compat=newest}

\usepackage{ifthen}
\graphicspath{{./figs/}}

\newcommand{\N}{\mathbb{N}}
\begin{document}

\title{Time-aware Test Case Execution Scheduling for Cyber-Physical Systems}

\author{Morten Mossige \inst{1,3}
  \and Arnaud Gotlieb\inst{2}
  \and Helge Spieker\inst{2}
  \and \\Hein Meling\inst{3}
  \and Mats Carlsson\inst{4}
}

\institute{
ABB Robotics, Bryne, Norway, \email{morten.mossige@uis.no}
\and
Simula Research Laboratory, Lysaker, Norway,
\email{\{arnaud,helge\}@simula.no}\thanks{These authors are supported by the Research
  Council of Norway (RCN) through the research-based innovation center Certus, under the SFI programme.}
\and
University of Stavanger, Stavanger, Norway, \email{hein.meling@uis.no}
\and
RISE SICS, Kista, Sweden, \email{mats.carlsson@ri.se}}

\maketitle

\newcommand{\includeBigFigures}      {true}
\newcommand{\insertEmptyFig}      {\missingfigure[figwidth=\textwidth]{Missig figure, Set includeBigFigures to true to include figure}}

\newcommand{\secref}[1]{Section~\ref{#1}}
\newcommand{\expref}[1]{Exp#1}
\newcommand{\figref}[1]{Figure~\ref{#1}}
\newcommand{\tabref}[1]{Table~\ref{#1}}
\newcommand{\algoref}[1]{Algorithm~\ref{#1}}
\newcommand{\listref}[1]{Listing~\ref{#1}}
\newcommand{\CI}{CI\xspace}
\newcommand{\SCR}{source control repository\xspace}
\newcommand{\TSU}{Test suite\xspace}
\newcommand{\rnd}{\textit{random}\xspace}
\newcommand{\naive}{\textit{greedy}\xspace}
\newcommand{\tvevery}{\textit{TV-everywhere}\xspace}

\newcommand{\TSM}{\textit{test suite minimisation}\xspace}
\newcommand{\TCP}{\textit{test case prioritization}\xspace}
\newcommand{\TCS}{\textit{test case scheduler}\xspace}
\newcommand{\RTT}{round-trip time\xspace}

\newcommand{\CAPSQUIZE}{\vspace{-1.3em}}

\newcommand{\EQSQUIZE}{\vspace{-0.75em}}

\newcommand{\MAKESPAN}{\ensuremath{C^*}\xspace}
\newcommand{\MSPANF}{\ensuremath{\MAKESPAN_{f}}\xspace}
\newcommand{\MSPANL}{\ensuremath{\MAKESPAN_{l}}\xspace}
\newcommand{\AVGMAKESPAN}{\ensuremath{\overline{C^*}}\xspace}
\newcommand{\TOTALTIME}{\ensuremath{T_t}\xspace}
\newcommand{\SOLVINGTIME}{\ensuremath{T_s}\xspace}
\newcommand{\TSNIC}{TC-Sched\xspace}
\newcommand{\RANDALGO}{\textit{R-SELECT}\xspace}
\newcommand{\BFALGO}{\textit{BF-SELECT}\xspace}
\newcommand{\Cumulatives}{\textsc{Cumulatives}\xspace}
\newcommand{\Disjunctive}{\textsc{Disjunctive}\xspace}
\newcommand{\Label}{\textsc{Label}}
\newcommand{\Minimize}{\textsc{Minimize}}
\newcommand{\CP}{Constraint Programming\xspace}
\newcommand{\cp}{constraint programming\xspace}
\newcommand{\model}{model\xspace}
\newcommand{\clpfd}{\texttt{clpfd}\xspace}
\newcommand{\NOTEST}      {\ensuremath{n}\xspace}
\newcommand{\NOMACH}      {\ensuremath{m}\xspace}
\newcommand{\NORE}        {\ensuremath{o}\xspace}
\newcommand{\TESTMINDUR}  {\ensuremath{d_{min}}\xspace}
\newcommand{\TESTMAXDUR}  {\ensuremath{d_{max}}\xspace}
\newcommand{\PTtoM}       {\ensuremath{p_{tm}}\xspace}
\newcommand{\PTtoR}       {\ensuremath{p_{tr}}\xspace}
\newcommand{\PTGtoM}      {\ensuremath{p_{tg}}\xspace}

\definecolor{l_gray}{rgb}{0.86,0.86,0.86}
\lstset{backgroundcolor=\color{l_gray},frame=none,basicstyle=\ttfamily\footnotesize}
\newcommand{\codelist}[3]{
  \r	enewcommand{\lstlistingname}{Listing}
  \lstset{basicstyle=\ttfamily\footnotesize, language=C, keepspaces=true, columns=fullflexible}
  \lstset{numbers=none,tabsize=2}
  \lstinputlisting[fontadjust, frame=tb, caption={#2}, label={#3}]{#1}
}

\definecolor{gray1}{gray}{0.8}
\definecolor{gray2}{gray}{0.85}
\definecolor{gray3}{gray}{0.75}
\definecolor{gray4}{gray}{0.65}

\newcommand{\drawschedule}[5]	
{
	begin{scope}
	{
		\draw[fill=#5] (#3,#2-\mshift) rectangle (#4+#3,#2+\task_h-\mshift);
		\node at (#3+#4*0.5,#2+0.5*\task_h-\mshift) {$#1$};
	}
}

\newcommand{\plotmaxspan}[1]{
	\begin{tikzpicture}
		\begin{axis}[
				title={#1},
				width=\linewidth,
				xlabel={Solving time},
				ylabel={Max span},
				mark size=0.0]
				\addplot table[] {data/defaultlabeling/#1};
				\addplot table[] {data/circularlabeling/#1};
				\legend{Default,Circular}
		\end{axis}
	\end{tikzpicture}
	\\
}

\makeatletter
\pgfplotsset{
    boxplot/hide outliers/.code={
        \def\pgfplotsplothandlerboxplot@outlier{}%
    }
}
\makeatother

\begin{abstract}
  Testing cyber-physical systems involves the execution of test cases on
  target-machines equipped with the latest release of a software control system.
  When testing industrial robots, it is common that the target machines need to
  share some common resources, e.g., costly hardware devices, and so there is a
  need to schedule test case execution on the target machines, accounting for
  these shared resources. With a large number of such tests executed on a
  regular basis, this scheduling becomes difficult to manage manually. In fact,
  with manual test execution planning and scheduling, some robots may remain
  unoccupied for long periods of time and some test cases may not be executed.

  This paper introduces \TSNIC, a time-aware method for automated test case
  execution scheduling. \TSNIC uses Constraint Programming to schedule
  tests to run on multiple machines constrained by the tests' access to shared
  resources, such as measurement or networking devices. The CP model is written
  in SICStus Prolog and uses the Cumulatives global constraint. Given a set of
  test cases, a set of machines, and a set of shared resources, \TSNIC produces
  an execution schedule where each test is executed once with minimal time
  between when a source code change is committed and the test results are reported to
  the developer. Experiments reveal that \TSNIC can schedule $500$ test cases
  over $100$ machines in less than $4$ minutes for $99.5\%$ of the instances. In
  addition, \TSNIC largely outperforms simpler methods based on a greedy
  algorithm and is suitable for deployment on industrial robot testing.
\end{abstract}

\section{Introduction}
\label{sec:intro}
\noindent
Continuous integration (CI) aims to uncover defects in early stages of software
development by frequently building, integrating, and testing software systems.
When applied to the development of cyber-physical
systems (CPS)\footnote{CPS can simply be seen as communicating
  embedded software systems.}, the process may include running integration test
cases involving real hardware components on different machines or machines
equipped with specific devices. In the last decade, CI has been recognized as an
effective process to improve software quality at reasonable
costs~\cite{duvall2007continuous,orso2014software,stolberg2009enabling,ERP14}.

Different from traditional testing methods, running a test case in CI requires
tight control over the \textit{round-trip time}, that is, the time from when a
source code change is committed until the success or failure of the build and
test processes is reported back to the developer~\cite{fowler2006continuous}.
Admittedly, the easiest way to minimize the round-trip time is simply to execute
as many tests as possible in the shortest amount of time. But the achievable
parallelism is limited by the availability of scarce global resources, such as a
costly measurement instrument or network device, and the compatible machines per
test case, targeting different machine architecture and operating systems. These
global resources are required in addition to the machine executing the test case
and thereby require parallel adjustments of the schedule for multiple machines.

Thus, computing an optimal test schedule with minimal round-trip time is a
challenging optimization problem. Since different test cases have different
execution times and may use different global resources that are locked during
execution, finding an optimal schedule manually is mostly impossible.
Nevertheless, manual scheduling still is state-of-the-practice in many
industrial applications, besides simple heuristics. In general, successful
approaches to scheduling use techniques from Constraint Programming (CP) and
Operations Research (OR), additionally metaheuristics are able to provide good
solutions to certain scheduling problems. We discuss these approaches further in
\secref{sec:existing_solutions}.

Informally, the optimal test scheduling problem (OTS) is to find an execution
order and assignment of all test cases to machines. Each test case has to be
executed once and no global resource can be used by two test cases at the same
time. The objective is to minimize the overall test scheduling and test
execution time. The assignment is constrained by the compatibility between test
cases and machines, that is, each test case can only be executed on a subset of
machines.

This paper introduces \TSNIC, a time-aware method to solve OTS. Using the
\Cumulatives~\cite{Aggoun199357,beldiceanu2002new} global constraint, we propose
a cost-effective constraint optimization search technique. This method allows us
to 1) automatically filter invalid test execution schedules, and 2) find among
possible valid schedules, those that minimize the global test execution time
(i.e., makespan). To the best of our knowledge, this is the first time the
problem of optimal scheduling test suite execution is formalized and a fully
automated solution is developed using constraint optimization techniques. \TSNIC
has been developed and deployed together with ABB Robotics, Norway.

An extensive experimental evaluation is conducted over test suites from
industrial software systems, namely an integrated control system for industrial
robots and a product line of video-conferencing systems. The primary goal in
this paper is to demonstrate the scalability of the proposed approach for CI
processes involving hundreds of test cases and tens of machines, which
corresponds to a realistic development environment. Furthermore, we demonstrate
the cost-effectiveness of integrating our approach within an actual CI process.

\section{Existing Solutions and Related Work}
\label{sec:existing_solutions}
\noindent
Automated solutions to address the OTS problem are not yet common practice. In
industrial settings, test engineers manually design the scheduling of test case
execution by allocating executions to certain machines at a given time or
following a given order. In practice, they manage the constraints as an
aggregate and try to find the best compromise in terms of the time needed to
execute the test cases. Keeping this process manual in \CI is paradoxical, since
every activity should, in principle, be automated.

Regression testing \cite{OSH04}, i.e. the repeated testing of systems after changes were made, in \CI covers a broad area of research works, including automatic test case generation \cite{CAF14}, test suite prioritization and test suite reduction \cite{ERP14}.
There, the idea of controlling the time taken by optimization processes in test suite prioritization is not new \cite{DMT10}. 
In test suite prioritization, \cite{WSK06} proposed to use time-aware genetic algorithms to optimize the order in which to execute the test cases.
Zhang {\it et al.} further refined this approach in \cite{ZHG09} by using integer linear programming.
On-demand test suite reduction \cite{HZW12} also exploits integer linear programming for 
preserving the fault-detection capability of a test suite while performing test suite reduction. 
Cost-aware methods are also available for selecting minimal subsets of test cases covering a number of requirements \cite{LTK14,GM14}.
All these approaches participate in a general effort to better control the time allocated to the optimization algorithms when
they are used in \CI processes.
Note however that test suite execution scheduling is different to prioritization or reduction as it deals with the notion of scheduling in time the execution of all test cases, without paying attention to any prioritization or reduction.

Scheduling problems have been studied in other contexts for decades and an extensive body of research exists on resource-constrained approaches.
The scheduling domain is divided into distinct areas such as process execution scheduling in operating systems and scheduling of workforces in a
construction project. The scheduling problem of this paper belongs to a scheduling category named
resource-constrained project scheduling problem (RCPSP; see \cite{Brucker1999,Brucker2006,Hartmann2010} for an extensive overview).
RCPSP is concerned with finding schedules for resource-consuming tasks with precedence constraints in a fixed time horizon, such that the makespan is minimized \cite{Hartmann2010}.
From the angle of RCPSP, global resources can be expressed as \textit{renewable resources} which are available with exactly one unit per timestep and can therefore only be consumed by a single job per timestep.

RCPSP has been addressed by both exact methods \cite{Szeredi2016,kreter2015modeling,schutt2012maximising,Schutt2013}, as well as heuristic methods \cite{Hartmann2000,Kolisch2006}.
Due to the vast amount of literature, we will focus on CP/OR-methods most closely related to the work of this paper.
The clear trend in both CP and OR is to solve such problems with hybrid approaches, like, for instance, the work by Schutt et al.~\cite{Schutt2009} or Beck et al.~\cite{Beck2011}.
Furthermore, \textit{disjunctive scheduling problems}, a subfamily of RCPSP addressing unary resources (in our terms global resources), have been effectively solved, e.g. by lazy clause generation~\cite{Siala2015}.

RCPSP is considered to be a generalization of \textit{machine scheduling problems} where \textit{job shop scheduling} (JSS) is one of the best known~\cite{herroelen1998resource}.
JSS is the special case of RCPSP where each operation uses exactly one resource, and FJSS (\textit{flexible job shop scheduling}) further extends JSS such that each operation can be processed on any machine from a given set.
The FJSS is known to be NP-hard~\cite{behnke2012test}.

While OTS is closely related to FJSS, and efficient approaches to FJSS are known \cite{brandimarte1993routing,schutt2013scheduling}, there are some differences.
First, in OTS, execution times are machine-independent.
Second, each job in OTS consists of only one operation, while in FJSS one job can contain several operations, where there are precedences between the operations.
Finally, some operations additionally require exclusive access to a global resource, preventing overlap with other operations.

\section{Problem Modeling}
\label{sec:notation_and_background}
\noindent
This section contains a formal definition of the OTS problem for test suite
execution on multiple machines with resource constraints. Based on this
definition, we propose a constraint optimization model using \Cumulatives global
constraint.

\subsection{Optimal Test Case Execution Scheduling}
\label{sec:optimal_scheduling}
\noindent {\it Optimal test case scheduling}\footnote{OTS was part of the
  Industrial Modelling Competition at CP 2015.} (OTS) is an optimization problem
$(\mathcal{T},\mathcal{G},\mathcal{M},d,g,f)$, where $\mathcal{T}$ is a set of
$\NOTEST$ test cases along with a function $d:\mathcal{T} \longrightarrow \N$
giving each test case a duration $d_i$; a set of global resources $\mathcal{G}$
along with a function $g:\mathcal{T} \longrightarrow 2^{\mathcal{G}}$ that
describes which resources are used by each test case; and a set of machines
$\mathcal{M}$ and a function $f: \mathcal{T} \longrightarrow 2^{\mathcal{M}}$
that assigns to each test case a subset of machines on which the test case can
be executed. The function $d$ is usually obtained by measuring the execution
time of each test case in previous test campaigns and by over-approximating each
duration to account for small variations between the different execution
machines.
OTS is the optimization problem of finding an execution ordering and assignment
of all test cases to machines, such that each test case is executed once, no
global resource is used by two test cases at the same time, and the overall test
execution time, \TOTALTIME, is minimized. We define \TOTALTIME as the
time needed to compute the schedule (\SOLVINGTIME ) plus the time needed to execute
the schedule (\MAKESPAN), $\TOTALTIME = \SOLVINGTIME + \MAKESPAN$.
Machine assignment and test case execution ordering can be described either by a
time-discretized table containing a line per machine or a starting time for each
test case and its assignment to a given machine.

The problem addressed in this paper aims to execute each test case once while
minimizing the total duration of the execution of the test cases. That is, to
find an assignment $a: \mathcal{T} \longrightarrow \mathcal{M}$ and an execution
order for each machine to run its test cases.

In its basic version, the OTS problem includes the following constraints:

\noindent\textbf{Disjunctive scheduling:} Two test cases cannot be
executed at the same time on a single machine.\\
\textbf{Non-preemptive scheduling:} The execution of a test case cannot be
temporarily interrupted to execute another test case on the same machine.\\
\textbf{Non-shared resources:} When a test case uses a global resource, no other
test case needing this resource can be executed at the same time.\\
\textbf{Machine-independent execution time:} The execution time of a test case
is assumed to be independent of the executing machine. This is reasonable for
test cases in which the time is dominated by external physical factors such as a
robot's motion, the opening of a valve, or sending an Ethernet frame. Such test
cases typically have execution times that are uncorrelated with machine
performance. In any case, a sufficient over-approximation will satisfy the
assumption.

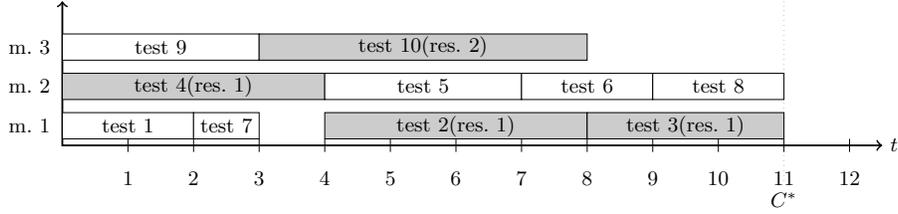
\begin{figure}[t]
\centering

\resizebox{\columnwidth}{!}
{
	\begin{tikzpicture}
	
	\def\task_h{0.4}
	\def\max_t{12}
	\def\m_ma{2.2}
	\def\mshift{0.5}

	\def\mone{0.6}
	\def\mtwo{1.2}
	\def\mthree{1.8}

	\draw [<->,thick] (0,\m_ma) node (yaxis) [above] {} |- (\max_t+0.5,0) node (xaxis) [right] {$t$};      
	
	\foreach \i in {1,2,...,\max_t} 
	{
		\node at (\i,-0.5) {\i};
		\draw[ultra thin] (\i, -0.1) -- (\i, 0.1);
	}

	\node at (-0.5, \mone+0.5*\task_h-\mshift) {$\text{m.~} 1$};
	\node at (-0.5, \mtwo+0.5*\task_h-\mshift) {$\text{m.~} 2$};
	\node at (-0.5, \mthree+0.5*\task_h-\mshift) {$\text{m.~} 3$};

	\drawschedule{\text{test~} 1}{\mone}{0}{2}{white}
	\drawschedule{\text{test~} 7}{\mone}{2}{1}{white}
	\drawschedule{\text{test~} 2(\text{res.~} 1)}{\mone}{4}{4}{gray1}
	\drawschedule{\text{test~} 3(\text{res.~} 1)}{\mone}{8}{3}{gray1}

	\drawschedule{\text{test~} 4(\text{res.~} 1)}{\mtwo}{0}{4}{gray1}
	\drawschedule{\text{test~} 5}{\mtwo}{4}{3}{white}
	\drawschedule{\text{test~} 6}{\mtwo}{7}{2}{white}
	\drawschedule{\text{test~} 8}{\mtwo}{9}{2}{white}

	\drawschedule{\text{test~} 9} {\mthree}{0}{3}{white}
	\drawschedule{\text{test~} 10(\text{res.~} 2)}{\mthree}{3}{5}{gray1}

	\draw[ultra thin, dotted] (11, \m_ma) -- (11, -0.6) node (cmax) [below] {$\MAKESPAN$} ;

	\end{tikzpicture}
}

 \caption{An optimal solution to the scheduling problem given in \tabref{tab:test_suite_ex1}.
Test cases in light gray require exclusive access to a global resource.}
\label{fig:task_sched1}
\CAPSQUIZE
\end{figure}

\noindent
There are cases where OTS can be trivially solved, e.g. with only one machine
executing all test cases in sequence. Indeed, the global execution time remains
unchanged, whatever the execution order. Similarly, when there are no global
resources and when test cases can be executed on any available machine, then
simply allocating the longest test cases first to the available execution
machine easily calculates a best-effort solution.
\noindent
\textbf{Example} Considering the test suite in \tabref{tab:test_suite_ex1}, we
present a small example. %
Let $\mathcal{T}$ be the test cases $\{1,\ldots,10\}$, $\mathcal{G}$ be the
global resources $\{1,2\}$, and $\mathcal{M}$ be the machines $\{1,2,3\}$. The
machines on which each test case in $\mathcal{T}$ can run is given in
\tabref{tab:test_suite_ex1}. This table can be extracted by analyzing the test
scripts or querying the test management. By sharing the same resource $1$, test
cases $2,3,4$ cannot be executed at the same time, even if their execution is
scheduled on different machines. Since test case $7$ can only be executed on
machine $1$, test case $8$ on machine $2$, test case $9$ on machine $3$, and
test case $10$ on machines $1$ or $3$, we have to solve a complex scheduling
problem. One possible \textit{optimal} schedule is given in
\figref{fig:task_sched1}, where the time needed to execute the test campaign is
$\MAKESPAN=11$. For this small problem the solving time, \SOLVINGTIME, can be
assumed to be very short, so the total execution time will be $\TOTALTIME
\approx \MAKESPAN$.

\begin{table}[t]
  \small
	\setlength{\tabcolsep}{1em}
	\renewcommand{\arraystretch}{1.0} 
	\centering
	\begin{threeparttable}[b]
		\caption{Test suite for example.}
		\label{tab:test_suite_ex1}
		
		\begin{tabular}{@{}cccc@{}}
		\toprule
		Test &  Duration & Executable on &  Use of global resource\\ 
		\midrule 
		$1$    & 2 & $1,2,3$  & -     \\
		$2$    & 4 & $1,2,3$  & $1$ \\
		$3$    & 3 & $1,2,3$  & $1$ \\
		$4$    & 4 & $1,2,3$  & $1$ \\
		$5$    & 3 & $1,2,3$  & -     \\
		$6$    & 2 & $1,2,3$  & -     \\
		$7$    & 1 & $1$          & -     \\
		$8$    & 2 & $2$          & -     \\
		$9$    & 3 & $3$          & -     \\
		$10$   & 5 & $1,3$      & $2$     \\
		\bottomrule
		\end{tabular}

  \end{threeparttable}
  \CAPSQUIZE
 \end{table} %

\subsection{The \Cumulatives Global Constraint}
\label{sec:cum_constraint}
\noindent
The \Cumulatives global constraint \cite{beldiceanu2002new} is a powerful
tool for modeling cumulative scheduling of multiple operations on multiple
machines, where each operation can be set up to consume a given amount
of a resources, and each machine can be set up to provide a
given amount of resources.

$\Cumulatives([O_1,\ldots,O_n], [c_1,\ldots,c_p])$\footnote{In
  \cite{beldiceanu2002new} an additional third argument to \Cumulatives, $Op \in
  \{\leq, \geq\}$ is defined. We omit it throughout our work and always set $Op
  =\,\leq$.} constrains $n$ operations on $p$ machines such that the total
resource consumption on each machine $j$ does not exceed the given threshold
$c_j$ at any time \cite{Carlsson:1997}. An operation $O_i$ is typically
represented by a tuple $(S_i, d_i, E_i, r_i, M_i)$\footnote{Throughout the
  paper, lower-case characters are used to represent constants and upper-case
  characters are used to represent variables.} where $S_i$ (resp. $E_i$) is a
variable that denotes the starting (resp. ending) instant of the operation,
$d_i$ is a constant representing the total duration of the operation, $r_i$ is a
constant representing the amount of resource used by the operation. $S_i, E_i$
and $M_i$ are bounded integer variables. $S_i$ and $E_i$ have the domains $est_i
\ldots let_i$, where $est_i$ denotes the operation's earliest starting time and
$let_i$ denotes its latest ending time and $let_i \geq est_i + d_i$. $M_i$ is
bounded by the number of machines available, that is $1,\ldots,p$. By reducing
the domain of $M_i$ it is possible to force a specific operation to be assigned
to only a subset of the available machines, or even to one specific machine. It
is worth noting that this formalization implicitly uses discrete time instants.
Indeed, since $est_i$ and $let_i$ are integers, a function associating each time
instant to the current executed operations can automatically be constructed.
Formally, if $h$ represents an instant in time, we have: \EQSQUIZE
\[ r_i^h = \left\{ 
	\begin{array}{l l}
 		r_i & \quad \text{if $S_i \leq h < S_i + d_i$}\\
		0   & \quad \text{otherwise}
	\end{array} \right.
\]
\EQSQUIZE

\noindent
$\Cumulatives$ holds if and only if, for every operation $O_i$, $S_i + d_i =
E_i$, and, for all machines $k$ and instants $h$, $\sum_{i \mid M_i=k} r^h_i \leq
c_k$. In fact, $\Cumulatives$ captures a disjunctive relation between different
scenarios and applies deductive reasoning to the possible values in the domains
of its variables. This constraint provides a cost-effective process for pruning
the search space of some impossible schedules.

\subsection{Modeling Test Case Execution Scheduling}
\label{sec:use_cum_constraint}
\noindent
This section shows how the \Cumulatives constraint can be used to model a
schedule. In this small example, we disregard the use of global resources, and
the constraints that some operations can only be executed on a subset of the
available machines, since that will be covered in
\secref{sec:use_cum_constraint_global}. By the schedule in
\figref{fig:task_sched1}, we have ten operations
$\mathcal{O}=\{O_1,\ldots,O_{10}\}$ and three available machines. By encoding
the data from \tabref{tab:test_suite_ex1}, we get $O_1=(S_1,2,E_1, 1, M_1)$,
$O_2=(S_2,4,E_2, 1, M_2)\ldots$, $O_{10}=(S_{10},5,E_{10}, 1, M_{10})$, $c_1=1$,
$c_2=1$, $c_3=1$. Note that each operation has a resource consumption of one and
all three machines have a resource capacity of one. This implies that one
machine can only execute one operation at a time. Here, a resource refers to an
execution machine and not to a global resource.

\subsection{Introducing Global Resources}
\label{sec:use_cum_constraint_global}
\noindent
As mentioned above, global resources corresponding to physical equipment such as
valves, air sensors, measurement instruments, or network devices, have limited
and exclusive access. To avoid concurrent access from two test cases, additional
constraints are introduced. Note that global resources must not be confused with
the resource consumption or resource bounds of operations and machines.

The \Cumulatives constraint does not support native modelling of these global
resources without additional, user-defined constraints. However, there are ways
to model exclusive access to such global resources by means of further
constraints. The naive approach to prevent two operations from overlapping is to
consider constraints over the start and stop time of the operations. For
instance, if $O_1$ and $O_2$ both require exclusive access to a global resource,
then the constraint $E_1 \le S_2 \lor E_2 \le S_1$ can be added. A less naive
approach is to use a $\Disjunctive(\mathcal{O}^k)$ constraint per global
resource $k$, where $\mathcal{O}^k$ is the set of tasks that require that global
resource, and \Disjunctive prevents any pair of tasks from
overlapping.

Referring to the example in \figref{fig:task_sched1}, there are ten operations
to be scheduled on three machines, and two global resources, $1$ and $2$. The
basic scheduling constraint is set up as explained in
\secref{sec:use_cum_constraint}. Yet another way to model the global resources
is to treat each resource as a new quasi-machine $1'$ corresponding to
$c_{1'}=1$ and $2'$ corresponding to $c_{2'}=1$. For each operation requiring a
global resource, we create a ``mirrored'' operation of the corresponding
quasi-machine: $\mathcal{O}'_{1}=\{O'_2,O'_3,O'_4\}$ and
$\mathcal{O}'_{2}=\{O'_{10}\}$. Finally, we can express the schedule with a
single constraint: $\Cumulatives(\mathcal{O} \cup \mathcal{O}'_{1} \cup
\mathcal{O}'_{2},[c_1,c_2,c_3,c_{1'}, c_{2'}])$. For each operation in
$\mathcal{O}'_{1}$ and $\mathcal{O}'_{2}$ we also reuse the same domain
variables for start-time, duration and end-time. The operation $O_4$ will be
forced to have the same start-/end-time as $O'_4$, while they are scheduled on
two different machines $2$ and $1'$.

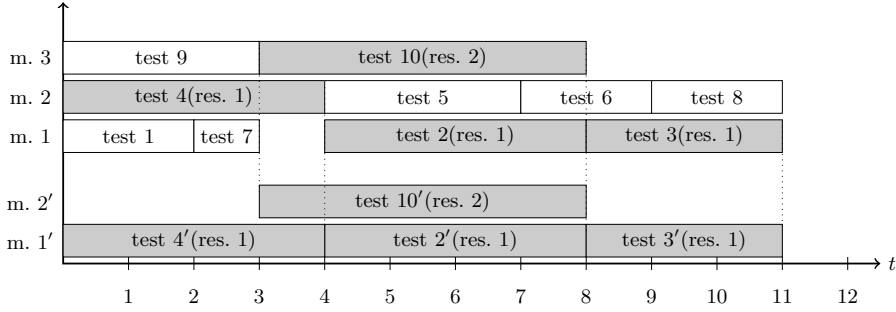
\begin{figure}[t]

\resizebox{\columnwidth}{!}
{
	\begin{tikzpicture}
	
	\def\task_h{0.5}
	\def\max_t{12}
	\def\m_ma{4}
	\def\mshift{0.5}
	
	\def\mone{2.2}
	\def\mtwo{2.8}
	\def\mthree{3.4}
	\def\mrone{0.6}
	\def\mrtwo{1.2}
	\def\loffs{-0.1}

	\draw [<->,thick] (0,\m_ma) node (yaxis) [above] {} |- (\max_t+0.5,0) node (xaxis) [right] {$t$};      
	
	\foreach \i in {1,2,...,\max_t} 
	{
		\node at (\i,-0.5) {\i};
		\draw[ultra thin] (\i, -0.1) -- (\i, 0.1);
	}

	\node at (-0.5, \mone+0.5*\task_h-\mshift) {$\text{m.~}1$};
	\node at (-0.5, \mtwo+0.5*\task_h-\mshift) {$\text{m.~}2$};
	\node at (-0.5, \mthree+0.5*\task_h-\mshift) {$\text{m.~}3$};
	
	\node at (-0.5, \mrone+0.5*\task_h-\mshift) {$\text{m.~}1'$};
	\node at (-0.5, \mrtwo+0.5*\task_h-\mshift) {$\text{m.~}2'$};

	\drawschedule{\text{test~}1}{\mone}{0}{2}{white}
	\drawschedule{\text{test~}7}{\mone}{2}{1}{white}
	\drawschedule{\text{test~}2(\text{res.~}1)}{\mone}{4}{4}{gray1}
	\drawschedule{\text{test~}3(\text{res.~}1)}{\mone}{8}{3}{gray1}

	\drawschedule{\text{test~}4(\text{res.~}1)}{\mtwo}{0}{4}{gray1}
	\drawschedule{\text{test~}5}{\mtwo}{4}{3}{white}
	\drawschedule{\text{test~}6}{\mtwo}{7}{2}{white}
	\drawschedule{\text{test~}8}{\mtwo}{9}{2}{white}

	\drawschedule{\text{test~}9} {\mthree}{0}{3}{white}
	\drawschedule{\text{test~}10(\text{res.~}2)}{\mthree}{3}{5}{gray1}

	\drawschedule{\text{test~}2'(\text{res.~}1)}{\mrone}{4}{4}{gray1}
	\drawschedule{\text{test~}3'(\text{res.~}1)}{\mrone}{8}{3}{gray1}
	\drawschedule{\text{test~}4'(\text{res.~}1)}{\mrone}{0}{4}{gray1}

	\drawschedule{\text{test~}10'(\text{res.~}2)}{\mrtwo}{3}{5}{gray1}

	\draw[thin, dotted] (3, \mthree+\loffs) -- (3, \mrtwo+\loffs);
	\draw[thin, dotted] (8, \mthree+\loffs) -- (8, \mrone+\loffs);
	\draw[thin, dotted] (4, \mtwo+\loffs)   -- (4, \mrone+\loffs);
	\draw[thin, dotted] (11, \mone+\loffs)  -- (11, \mrone+\loffs);

	\end{tikzpicture}
}

 	\label{fig:globresource}
	\caption{Modeling global resources by creating quasi-machines and \Cumulatives}
	\CAPSQUIZE
\end{figure}

\section{The \TSNIC Method}
\label{sec:approach}
\noindent
This section describes our method, \TSNIC, to solve the OTS problem. It is a
{\it time-constrained cumulative scheduling technique}, as 1)~it allows to keep
fine-grained control over the time allocated to the constraint solving process
(i.e., {\it time-constrained}), 2)~it encodes exclusive resource use with
constraints (i.e., {\it constraint-based}), and 3)~it solves the problem by
using the \Cumulatives\ constraint. The \TSNIC method is composed of three
elements, namely, the constraint model described in \secref{Sec:TSNIC1}, the
search procedure described in \secref{Sec:TSNIC2}, and the time-constrained
minimization process described in \secref{Sec:TSNIC3}.

\subsection{Constraint Model}
\label{Sec:TSNIC1}
\noindent
We encode the OTS problem with one $\Cumulatives(\mathcal{O},\mathcal{C})$
constraint, one $\Disjunctive(\mathcal{O}^k)$ constraint per global resource
$k$, using the second scheme from \secref{sec:use_cum_constraint_global}, and a
search procedure able to find an optimal schedule among many feasible schedules.
Each test case $i$ is encoded as an operation $(S_i, d_i, E_i, 1, M_i)$ as
explained in \secref{sec:cum_constraint}. $\mathcal{O}$ is simply the array of
all such operations and $\mathcal{C}$ is an array of $1$s of length equal to the
number of machines.
Suppose that there are three execution machines numbered $1,2$, and $3$; then,
to say that test $i$ can be executed on any machine, we just add the domain
constraint $M_i \in \{1,2,3\}$, whereas to say that test $i$ can only be
executed on machine $1$, we replace $M_i$ by $1$.
Finally, to complete the model, we introduce the variable $\mathit{MakeSpan}$
representing the completion time of the entire schedule and seek to minimize it.
$\mathit{MakeSpan}$ is lower bounded by the ending time of each individual test
case. The generic model is captured by:
\begin{equation}\label{eq:model}
  \begin{array}{l}
  \Cumulatives({\mathcal{O}},{\mathcal{C}})\,\land \\
  \forall\,\text{global resource } k : \Disjunctive(\mathcal{O}^k)\,\land \\
  \forall\,1 \leq i \leq n : M_i \in f(i)\,\land\\
  \forall\,1 \leq i \leq n : E_i \leq \mathit{MakeSpan}\,\land \\
  \Label(\Minimize(\mathit{MakeSpan}), [S_1, M_1, \ldots, S_n, M_n])
  \end{array}
\end{equation}
\noindent
Note that the ending times depend functionally on the starting times. Thus, a
solution to the OTS problem can be obtained by searching among the starting
times and the assignment of test cases to execution machines.

\subsection{Search Procedure}
\label{Sec:TSNIC2}
\noindent
Our search procedure is called \textit{test case duration splitting}, and is a
branch-and-bound search that seeks to minimize the $\mathit{Makespan}$. The
procedure makes two passes over the set of test cases. A key idea is to allocate
the most demanding test cases first. To this end, the test cases are initially
sorted by decreasing $r_i$ where $r_i$ is the number of global resources used by
test case $i$, breaking ties by choosing the test case with the longest
duration $d_i$.

In Phase~1, two actions are performed on each test case. First, in order to
avoid a large branching factor in the choice of start time and to effectively
fix the relative order among the tasks on the same machine or resource, we split
the domain of the start variable, forcing an obligatory part of the
corresponding task, as described in~\cite[Section~3.6]{simonis2008search}. Next,
in order to balance the load on the machines, we choose machines in round-robin
fashion. These two choices are of course backtrackable, to ensure completeness
of the search procedure.

Note that at the end of Phase~1, the constraint system effectively forms a
directed acyclic graph where every node is a task and every arc is a precedence
constraint induced by the relative order. It is well known that such constraint
systems can be solved without search by topologically sorting the start
variables and assigning each of them to its minimal value. This is Phase~2 of
the search.

In this procedure, the load-balancing component has shown to be particularly
effective in a CI context and makes the first solution found a good compromise
between solving and execution time of the schedule, which is one of the key
factors in CI. Our preliminary experiments concluded, that the presented
strategy provided the best compromise between cost and solution quality.
Furthermore, we tried a more precise but costlier load-balancing scheme, but it
did not significantly improve the quality. We also tried to sort the tests by
decreasing $d_i \cdot (r_i+1)$, which did not significantly improve the quality,
either.

\subsection{Time-constrained Minimization}
\label{Sec:TSNIC3}
\noindent
The third necessary ingredient of the \TSNIC method is to perform branch-and-bound search under a time contract.
That is, to settle on the schedule with the shortest $\mathit{MakeSpan}$ found when the time contract ends.
When the number of test cases grows to be several hundred, finding a globally optimal schedule may become an intractable problem\footnote{The general cumulative scheduling problem
  is known to be NP-hard \cite{baptiste2001constraint}.}, but in practical applications it is often sufficient to find a ``best-effort'' solution.
This leads to the important question to select the most appropriate contract of time for the minimization process, as the time used to optimize the schedule is not available to actually execute the schedule.
We address this question in the experimental evaluation.

\section{Implementation and Exploitation}
\label{sec:implementation}
\noindent
This section details our implementation of the \TSNIC method and its insertion into \CI.
We implemented the \TSNIC method in SICStus Prolog\cite{carlssonsicstus}. 
The \Cumulatives constraint is available as part of the \clpfd library \cite{Carlsson:1997}.
The \clpfd library also provides an implementation of the time-constrained branch-and-bound with the option to express individual search strategy (see \secref{Sec:TSNIC2}).
Using \clpfd, a generic constraint model for the \TSNIC method is designed, which takes an OTS problem as input and returns an (quasi-)optimal schedule.

Since \TSNIC is designed to run as part of a \CI process, we describe how it can be integrated within the \CI environment. 
Because \CI environments change and test cases and agents are constantly added or removed, \TSNIC has to be provided with a list of test cases and available machines at runtime.
Furthermore, an estimation of the test case durations on the available agents has to be provided.
This can either be gathered from historical execution data and then (over-)estimated to account for differences in execution machines, or, for some kinds to test suites, they are fixed and can be precisely given \cite{MGH14a}, e.g. for robotic applications where the duration is determined by the movement of the robot.

 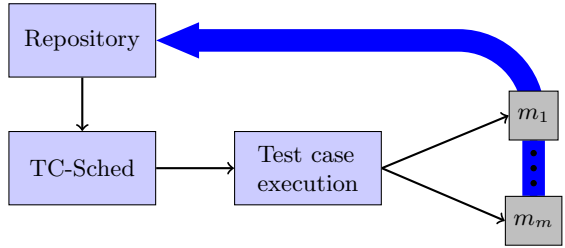
\begin{figure}[t]
 	\centering

\tikzstyle{block} = [draw, fill=blue!20, rectangle, minimum height=3em, minimum width=6em]
\tikzstyle{machine} = [draw, fill=gray!50, rectangle, minimum height=2em, minimum width=2em]
\tikzstyle{adot} = [draw, fill=black, circle, inner sep=0pt,minimum size=2pt]
\tikzstyle{myarrows}=[smooth, line width=1mm,draw=blue,-triangle 45,postaction={draw, line width=3mm, shorten >=4mm, -}]

\begin{tikzpicture}[->]

	\node [block, align=left, name=tcsched] (tcsched) {\TSNIC};
	\node [block, align=left, name=execute, right of=tcsched,node distance=3cm ] (execute) {Test case \\execution};
	\node [block, align=left, name=repo, above of=tcsched,node distance=1.7cm ] (repo) {Repository};
	\node [adot, align=left, name=adotc, right of=execute,node distance=3cm ] (adotc) {};
	\node [adot, align=left, name=adoto, above of=adotc,node distance=0.2cm ] (adoto) {};
	\node [adot, align=left, name=adotu, below of=adotc,node distance=0.2cm ] (adotu) {};
	\node [machine, align=left, name=mone, above of=adoto,node distance=0.5cm ] (mone) {$m_1$};
	\node [machine, align=left, name=mn, below of=adotu,node distance=0.5cm ] (mn) {$m_{\NOMACH}$};

	\draw [thick](repo.south)->(tcsched.north);
	\draw [thick](tcsched.east)->(execute.west);
	\draw [thick](execute.east)->(mone.west);
	\draw [thick](execute.east)->(mn.west);

	\begin{scope}[on background layer]
		\draw[myarrows] (mn) -- (mone) arc(0:90:1) -- (repo);
	\end{scope}

\end{tikzpicture}  	%
 	\caption{Integration of \TSNIC into a \CI process. The test case schedule solved
 	by \TSNIC is transmitted for execution to the machines in the machine
 	pool, $\cal M$. The results including actual test case durations are then 
 	feed back into the repository.}
 	\label{fig:systemoverview}
 	\CAPSQUIZE
 \end{figure}

 A test campaign in a \CI cycle is typically initiated upon a successful build
 of the software being tested. As a first step, all machines available for test
 execution are identified and updated with the newly built software. Then,
 \TSNIC takes as input the test cases of the test campaign and the previous test
 case execution times from the storage repository. After \TSNIC calculated an
 optimal schedule, that schedule is handed over to a dedicated dispatch server
 which is responsible for distributing the test cases to the physical machines
 and the actual execution. Finally, after the test execution finished, the
 overall result of the test campaign is reported back to the users and the
 storage repository is updated with the latest test case execution times. Of
 course, minimizing the {\it round-trip time} leads to earlier notifications of
 the developers in case the software system fails and helps to improve the
 development cycle in CI.

\section{Experimental Evaluation}
\label{sec:exp_evaluation}
\noindent
This section presents our findings from the experimental evaluation of \TSNIC.
To this end, we address the following three research questions:

\noindent{\bf RQ1:} How does the first solution provided by \TSNIC compare with
	simpler scheduling methods in terms of schedule execution time? This research question states the crucial question 
        of whether using complex constraint optimization is useful despite simpler approaches being available
        at almost no cost to implement.\\
{\bf RQ2:} For \TSNIC, will an increased investment in the solving time in \TSNIC
	reduce the overall time of a \CI cycle? This question is about finding the
	most appropriate trade-off between the solving time and the execution time of
	the test campaign in the proposed approach.\\
{\bf RQ3:} In addition to random OTS problem instances, can \TSNIC efficiently and effectively handle 
         industrial case studies? These cases can lead to structured problems which exhibit very different properties than
         random instances.

All experiments were performed on a 2.7~GHz Intel
Core~i7 processor with 16~GB RAM, running SICStus Prolog 4.3.5 on a Linux
operating system.

\subsection{Experimental Artifacts}
\label{sec:randomts}
\noindent
To answer RQ1, we implemented two scheduling methods, referred to as the
\rnd method and the \naive method.

The \rnd method works as follows: It first picks a test case at random and then
picks a machine at random such that no resource constraint is violated. Finally,
the test case is assigned the lowest possible starting time on the selected
machine. The \naive method is more advanced. At first, it assigns test cases by
decreasing resource demands. Afterwards, test cases without any resource demands
are assigned to the remaining machines. For each assignment, the machine that
can provide the earliest starting time is selected. Note that none of the two
methods can backtrack to improve upon the initial solution.

The reason we have chosen to compare with these two methods is threefold: 1) As
explained in \secref{sec:existing_solutions}, we are not aware of any previously
published work related to test case execution scheduling, which means that there
is no baseline to compare against; 2) From cooperation with our industrial
partners, we know that this is, in the best case, the industrial state of the
art (i.e., non-optimal schedules computed manually); 3) We manually checked the
results on simple schedules and found them to be satisfactory, so they are a
suitable comparison.

To answer our research questions, we have considered randomly generated benchmarks and industrial case studies.
Although there are benchmark test suites for both JSS and FJSS, e.g., \cite{taillard1993benchmarks} or \cite{behnke2012test}, they cannot be used as a comparison baseline.
Furthermore, as our method approaches testing applications, a thorough evaluation on data from the target domain is justifiable.

We generated a benchmark library containing $840$ OTS instances\footnote{All
  generated instances are available in CSPLib, a library of test problems for
  constraint solvers \cite{csplib:prob073}}.
The library is structured by data collected from three different real-world test suites, provided by our
industrial partners: a test suite for video conferencing systems (VCS) \cite{MGS13}, a test suite for integrated painting systems (IPS) \cite{MGH14a},
and a test suite for a mobile application called \tvevery. 
VCS is a test suite for testing commercial video conferencing systems, developed by CISCO Systems, Norway.
It contains $132$ test cases and $74$ machines. The duration of test cases 
varies from 13~seconds to 4~hours, where the vast majority has a duration between 100\,s and 800\,s.
The IPS test suite aims at testing a distributed paint control system for complex industrial robots,
developed at ABB Robotics, Norway.
It contains $33$ test cases, with duration ranging from 1\,s to
780\,s, and $16$ distinct machines.
There are two global resources for this test suite, an airflow meter and a simulator for an optical encoder.
\tvevery is a mobile application that allows users to watch TV on tablets, smart
phones, and laptops. Its test suite only contains manual test cases, but, in our benchmark,
it serves as a useful example of a test suite with a large number of constraints
limiting the number of possible machines for each test case.

Based on data from the three industrial test suites, we composed $14$ groups of test suites, denoted TS1-TS14, with randomized assignments of test cases to machines and exclusive usages of global resources.
Let $|T|$ be the number of test cases, and $|M|$ be the number of machines, and $|R|=\{3,5,10\}$ be the number of resources.
\tabref{tab:random_ts_overview1} gives an overview of the groups of test suites. 
For test suite TS$x$, we write TS$x$R3, TS$x$R5, or TS$x$R10 to indicate the number of resources.

\begin{table}[t]
	\small
	\setlength{\tabcolsep}{1em}
	\renewcommand{\arraystretch}{1.1} 
	\centering
	\begin{threeparttable}[b]
		\caption{Randomly generated test suites.}
		\label{tab:random_ts_overview1}
		
		\begin{tabular}{@{}cc|cccccc@{}}
                  \toprule

		\multicolumn{2}{c}{\# of tests} & 20  & 30  & 40  & 50  & 100  & 500 \\ 
		\midrule 
		\multirow{4}{*}{ \rotatebox[origin=c]{90}{\# machines}} 
		& 100  &  -  &  -  &  -  &  -  &  -   & TS11 \\ 
		& 50   &  -  &  -  &  -  &  -  & TS8  & TS12\\ 
		& 20   &  -  & TS2 & TS4 & TS6 & TS9  & TS13 \\ 
		& 10   & TS1 & TS3 & TS5 & TS7 & TS10 & TS14 \\ 

		\bottomrule
		\end{tabular}

  \end{threeparttable}
  \CAPSQUIZE
 \end{table}

For each of the $14\cdot3$ variants, we generated $20$ random test suites.
The duration of each test case was chosen randomly between $1$\,s and $800$\,s, and each test case had a 30\,\% chance of using a global resource. The number of resources was chosen randomly between $1$ and $|R|$.
A total of 80\,\% of the tests were
considered to be executable on all machines, while the remaining 20\,\% were
executable on a smaller subset of machines. For these tests, the number of
machines on which each test case could be executed was selected randomly between
1\,\% and 40\,\% of the number of available machines. This means that a test case
was executable either on all machines (part of the 80\,\% group) or only on at
most 40\,\% of the machines.
In total, we generated $14\cdot3\cdot20=840$ different test suites.

\subsection{RQ1: How does \TSNIC compare with simpler scheduling?}
\label{sec:rq1}
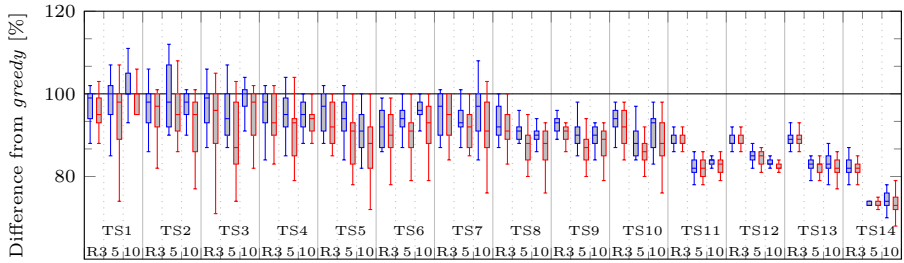
\begin{figure}[!t]
	\ifthenelse{\equal{\includeBigFigures}{true}}
	{

\begin{tikzpicture}
	\def\RND_COL{red}
	\def\FIRST_CP_COL{blue}
	\def\LAST_CP_COL{red}        
	\def\RSPACE{\,}
	
\begin{axis}[
  boxplot/draw direction=y,
  boxplot/variable width,
  boxplot/variable width min target=0.1,
  boxplot/every box/.style={fill=gray!50},
  width=1.02\textwidth,
  height=0.4\textwidth,
  xmin= 0,
  xmax=14,
  ymin=.6,
  ymax= 1.2,
  ytick={0.8,1,1.2},
  yticklabels={80, 100, 120},
  xmajorgrids,
  xminorgrids,
  minor tick num=2,
  major grid style={smooth},
  minor grid style={dotted},
  ylabel={\scriptsize Difference from \naive [\%]},
  ylabel near ticks,
  xtick={0,1,2,3,4,5,6,7,8,9,10,11,12,13,14},
  xticklabels={ { \tiny TS1 \\ \tiny R3\RSPACE{}5\RSPACE{}10 },    %
  							{ \tiny TS2 \\ \tiny R3\RSPACE{}5\RSPACE{}10 },    %
  							{ \tiny TS3 \\ \tiny R3\RSPACE{}5\RSPACE{}10 },    %
  							{ \tiny TS4 \\ \tiny R3\RSPACE{}5\RSPACE{}10 },    %
  							{ \tiny TS5 \\ \tiny R3\RSPACE{}5\RSPACE{}10 },    %
  							{ \tiny TS6 \\ \tiny R3\RSPACE{}5\RSPACE{}10 },    %
  							{ \tiny TS7 \\ \tiny R3\RSPACE{}5\RSPACE{}10 },    %
  							{ \tiny TS8 \\ \tiny R3\RSPACE{}5\RSPACE{}10 },    %
  							{ \tiny TS9 \\ \tiny R3\RSPACE{}5\RSPACE{}10 },    %
  							{ \tiny TS10\\ \tiny R3\RSPACE{}5\RSPACE{}10 },    %
  							{ \tiny TS11\\ \tiny R3\RSPACE{}5\RSPACE{}10 },    %
  							{ \tiny TS12\\ \tiny R3\RSPACE{}5\RSPACE{}10 },    %
  							{ \tiny TS13\\ \tiny R3\RSPACE{}5\RSPACE{}10 },    %
  							{ \tiny TS14\\ \tiny R3\RSPACE{}5\RSPACE{}10 } },  %
  x tick label style={ yshift = 1.7em  },
  x tick label as interval,
  xticklabel style={align=center},
  tick label style={font=\scriptsize}
  ]
                  
	\addplot[mark=none, black] coordinates { (0,1) (21.5,1) };

	\addplot[\FIRST_CP_COL, boxplot={hide outliers,draw position=0.10}] table[y=firstcp] {solvecompare_cums20/t20m10r3.dat};  
	\addplot[\LAST_CP_COL,  boxplot={hide outliers,draw position=0.25}] table[y=lastcp]  {solvecompare_cums20/t20m10r3.dat};  

	\addplot[\FIRST_CP_COL, boxplot={hide outliers,draw position=0.45}] table[y=firstcp]  {solvecompare_cums20/t20m10r5.dat};  
	\addplot[\LAST_CP_COL,  boxplot={hide outliers,draw position=0.60}] table[y=lastcp]   {solvecompare_cums20/t20m10r5.dat};  
	\addplot[\FIRST_CP_COL, boxplot={hide outliers,draw position=0.75}] table[y=firstcp] {solvecompare_cums20/t20m10r10.dat};  
	\addplot[\LAST_CP_COL,  boxplot={hide outliers,draw position=0.90}] table[y=lastcp]  {solvecompare_cums20/t20m10r10.dat};

	\addplot[\FIRST_CP_COL, boxplot={hide outliers,draw position=1.10}] table[y=firstcp] {solvecompare_cums20/t30m20r3.dat};  
	\addplot[\LAST_CP_COL,  boxplot={hide outliers,draw position=1.25}] table[y=lastcp]  {solvecompare_cums20/t30m20r3.dat};  

	\addplot[\FIRST_CP_COL, boxplot={hide outliers,draw position=1.45}] table[y=firstcp]  {solvecompare_cums20/t30m20r5.dat};  
	\addplot[\LAST_CP_COL,  boxplot={hide outliers,draw position=1.60}] table[y=lastcp]   {solvecompare_cums20/t30m20r5.dat};  

	\addplot[\FIRST_CP_COL, boxplot={hide outliers,draw position=1.75}] table[y=firstcp] {solvecompare_cums20/t30m20r10.dat};  
	\addplot[\LAST_CP_COL,  boxplot={hide outliers,draw position=1.90}] table[y=lastcp]  {solvecompare_cums20/t30m20r10.dat};

	\addplot[\FIRST_CP_COL, boxplot={hide outliers,draw position=2.10}] table[y=firstcp] {solvecompare_cums20/t30m10r3.dat};  
	\addplot[\LAST_CP_COL,  boxplot={hide outliers,draw position=2.25}] table[y=lastcp]  {solvecompare_cums20/t30m10r3.dat};  

	\addplot[\FIRST_CP_COL, boxplot={hide outliers,draw position=2.45}] table[y=firstcp]  {solvecompare_cums20/t30m10r5.dat};  
	\addplot[\LAST_CP_COL,  boxplot={hide outliers,draw position=2.60}] table[y=lastcp]   {solvecompare_cums20/t30m10r5.dat};  

	\addplot[\FIRST_CP_COL, boxplot={hide outliers,draw position=2.75}] table[y=firstcp] {solvecompare_cums20/t30m10r10.dat};  
	\addplot[\LAST_CP_COL,  boxplot={hide outliers,draw position=2.90}] table[y=lastcp]  {solvecompare_cums20/t30m10r10.dat};

	\addplot[\FIRST_CP_COL, boxplot={hide outliers,draw position=3.10}] table[y=firstcp] {solvecompare_cums20/t40m20r3.dat};  
	\addplot[\LAST_CP_COL,  boxplot={hide outliers,draw position=3.25}] table[y=lastcp]  {solvecompare_cums20/t40m20r3.dat};  

	\addplot[\FIRST_CP_COL, boxplot={hide outliers,draw position=3.45}] table[y=firstcp]  {solvecompare_cums20/t40m20r5.dat};  
	\addplot[\LAST_CP_COL,  boxplot={hide outliers,draw position=3.60}] table[y=lastcp]   {solvecompare_cums20/t40m20r5.dat};  

	\addplot[\FIRST_CP_COL, boxplot={hide outliers,draw position=3.75}] table[y=firstcp] {solvecompare_cums20/t40m20r10.dat};  
	\addplot[\LAST_CP_COL,  boxplot={hide outliers,draw position=3.90}] table[y=lastcp]  {solvecompare_cums20/t40m20r10.dat};

	\addplot[\FIRST_CP_COL, boxplot={hide outliers,draw position=4.10}] table[y=firstcp] {solvecompare_cums20/t40m10r3.dat};  
	\addplot[\LAST_CP_COL,  boxplot={hide outliers,draw position=4.25}] table[y=lastcp]  {solvecompare_cums20/t40m10r3.dat};  

	\addplot[\FIRST_CP_COL, boxplot={hide outliers,draw position=4.45}] table[y=firstcp]  {solvecompare_cums20/t40m10r5.dat};  
	\addplot[\LAST_CP_COL,  boxplot={hide outliers,draw position=4.60}] table[y=lastcp]   {solvecompare_cums20/t40m10r5.dat};  

	\addplot[\FIRST_CP_COL,   boxplot={hide outliers,draw position=4.75}] table[y=firstcp] {solvecompare_cums20/t40m10r10.dat};  
	\addplot[\LAST_CP_COL,  boxplot={hide outliers,draw position=4.90}] table[y=lastcp]  {solvecompare_cums20/t40m10r10.dat};

	\addplot[\FIRST_CP_COL,   boxplot={hide outliers,draw position=5.10}] table[y=firstcp] {solvecompare_cums20/t50m20r3.dat};  
	\addplot[\LAST_CP_COL,  boxplot={hide outliers,draw position=5.25}] table[y=lastcp]  {solvecompare_cums20/t50m20r3.dat};  

	\addplot[\FIRST_CP_COL,   boxplot={hide outliers,draw position=5.45}] table[y=firstcp]  {solvecompare_cums20/t50m20r5.dat};  
	\addplot[\LAST_CP_COL,  boxplot={hide outliers,draw position=5.60}] table[y=lastcp]   {solvecompare_cums20/t50m20r5.dat};  

	\addplot[\FIRST_CP_COL,   boxplot={hide outliers,draw position=5.75}] table[y=firstcp] {solvecompare_cums20/t50m20r10.dat};  
	\addplot[\LAST_CP_COL,  boxplot={hide outliers,draw position=5.90}] table[y=lastcp]  {solvecompare_cums20/t50m20r10.dat};

	\addplot[\FIRST_CP_COL,   boxplot={hide outliers,draw position=6.10}] table[y=firstcp] {solvecompare_cums20/t50m10r3.dat};  
	\addplot[\LAST_CP_COL,  boxplot={hide outliers,draw position=6.25}] table[y=lastcp]  {solvecompare_cums20/t50m10r3.dat};  

	\addplot[\FIRST_CP_COL,   boxplot={hide outliers,draw position=6.45}] table[y=firstcp]  {solvecompare_cums20/t50m10r5.dat};  
	\addplot[\LAST_CP_COL,  boxplot={hide outliers,draw position=6.60}] table[y=lastcp]   {solvecompare_cums20/t50m10r5.dat};  

	\addplot[\FIRST_CP_COL,   boxplot={hide outliers,draw position=6.75}] table[y=firstcp] {solvecompare_cums20/t50m10r10.dat};  
	\addplot[\LAST_CP_COL,  boxplot={hide outliers,draw position=6.90}] table[y=lastcp]  {solvecompare_cums20/t50m10r10.dat};

	\addplot[\FIRST_CP_COL,   boxplot={hide outliers,draw position=7.10}] table[y=firstcp] {solvecompare_cums20/t100m50r3.dat};  
	\addplot[\LAST_CP_COL,  boxplot={hide outliers,draw position=7.25}] table[y=lastcp]  {solvecompare_cums20/t100m50r3.dat};  

	\addplot[\FIRST_CP_COL,   boxplot={hide outliers,draw position=7.45}] table[y=firstcp]  {solvecompare_cums20/t100m50r5.dat};  
	\addplot[\LAST_CP_COL,  boxplot={hide outliers,draw position=7.60}] table[y=lastcp]   {solvecompare_cums20/t100m50r5.dat};  

	\addplot[\FIRST_CP_COL,   boxplot={hide outliers,draw position=7.75}] table[y=firstcp] {solvecompare_cums20/t100m50r10.dat};  
	\addplot[\LAST_CP_COL,  boxplot={hide outliers,draw position=7.90}] table[y=lastcp]  {solvecompare_cums20/t100m50r10.dat};

	\addplot[\FIRST_CP_COL,   boxplot={hide outliers,draw position=8.10}] table[y=firstcp] {solvecompare_cums20/t100m20r3.dat};  
	\addplot[\LAST_CP_COL,  boxplot={hide outliers,draw position=8.25}] table[y=lastcp]  {solvecompare_cums20/t100m20r3.dat};  

	\addplot[\FIRST_CP_COL,   boxplot={hide outliers,draw position=8.45}] table[y=firstcp]  {solvecompare_cums20/t100m20r5.dat};  
	\addplot[\LAST_CP_COL,  boxplot={hide outliers,draw position=8.60}] table[y=lastcp]   {solvecompare_cums20/t100m20r5.dat};  

	\addplot[\FIRST_CP_COL,   boxplot={hide outliers,draw position=8.75}] table[y=firstcp] {solvecompare_cums20/t100m20r10.dat};  
	\addplot[\LAST_CP_COL,  boxplot={hide outliers,draw position=8.90}] table[y=lastcp]  {solvecompare_cums20/t100m20r10.dat};

	\addplot[\FIRST_CP_COL,   boxplot={hide outliers,draw position=9.10}] table[y=firstcp] {solvecompare_cums20/t100m10r3.dat};  
	\addplot[\LAST_CP_COL,  boxplot={hide outliers,draw position=9.25}] table[y=lastcp]  {solvecompare_cums20/t100m10r3.dat};  

	\addplot[\FIRST_CP_COL,   boxplot={hide outliers,draw position=9.45}] table[y=firstcp]  {solvecompare_cums20/t100m10r5.dat};  
	\addplot[\LAST_CP_COL,  boxplot={hide outliers,draw position=9.60}] table[y=lastcp]   {solvecompare_cums20/t100m10r5.dat};  

	\addplot[\FIRST_CP_COL,   boxplot={hide outliers,draw position=9.75}] table[y=firstcp] {solvecompare_cums20/t100m10r10.dat};  
	\addplot[\LAST_CP_COL,  boxplot={hide outliers,draw position=9.90}] table[y=lastcp]  {solvecompare_cums20/t100m10r10.dat};

	\addplot[\FIRST_CP_COL,   boxplot={hide outliers,draw position=10.10}] table[y=firstcp] {solvecompare_cums20/t500m100r3.dat};  
	\addplot[\LAST_CP_COL,  boxplot={hide outliers,draw position=10.25}] table[y=lastcp]  {solvecompare_cums20/t500m100r3.dat};  

	\addplot[\FIRST_CP_COL,   boxplot={hide outliers,draw position=10.45}] table[y=firstcp]  {solvecompare_cums20/t500m100r5.dat};  
	\addplot[\LAST_CP_COL,  boxplot={hide outliers,draw position=10.60}] table[y=lastcp]   {solvecompare_cums20/t500m100r5.dat};  

	\addplot[\FIRST_CP_COL,   boxplot={hide outliers,draw position=10.75}] table[y=firstcp] {solvecompare_cums20/t500m100r10.dat};  
	\addplot[\LAST_CP_COL,  boxplot={hide outliers,draw position=10.90}] table[y=lastcp]  {solvecompare_cums20/t500m100r10.dat};

	\addplot[\FIRST_CP_COL,   boxplot={hide outliers,draw position=11.10}] table[y=firstcp] {solvecompare_cums20/t500m50r3.dat};  
	\addplot[\LAST_CP_COL,  boxplot={hide outliers,draw position=11.25}] table[y=lastcp]  {solvecompare_cums20/t500m50r3.dat};  

	\addplot[\FIRST_CP_COL,   boxplot={hide outliers,draw position=11.45}] table[y=firstcp]  {solvecompare_cums20/t500m50r5.dat};  
	\addplot[\LAST_CP_COL,  boxplot={hide outliers,draw position=11.60}] table[y=lastcp]   {solvecompare_cums20/t500m50r5.dat};  

	\addplot[\FIRST_CP_COL,   boxplot={hide outliers,draw position=11.75}] table[y=firstcp] {solvecompare_cums20/t500m50r10.dat};  
	\addplot[\LAST_CP_COL,  boxplot={hide outliers,draw position=11.90}] table[y=lastcp]  {solvecompare_cums20/t500m50r10.dat};

	\addplot[\FIRST_CP_COL,   boxplot={hide outliers,draw position=12.10}] table[y=firstcp] {solvecompare_cums20/t500m20r3.dat};  
	\addplot[\LAST_CP_COL,  boxplot={hide outliers,draw position=12.25}] table[y=lastcp]  {solvecompare_cums20/t500m20r3.dat};  

	\addplot[\FIRST_CP_COL,   boxplot={hide outliers,draw position=12.45}] table[y=firstcp]  {solvecompare_cums20/t500m20r5.dat};  
	\addplot[\LAST_CP_COL,  boxplot={hide outliers,draw position=12.60}] table[y=lastcp]   {solvecompare_cums20/t500m20r5.dat};  

	\addplot[\FIRST_CP_COL,   boxplot={hide outliers,draw position=12.75}] table[y=firstcp] {solvecompare_cums20/t500m20r10.dat};  
	\addplot[\LAST_CP_COL,  boxplot={hide outliers,draw position=12.90}] table[y=lastcp]  {solvecompare_cums20/t500m20r10.dat};

	\addplot[\FIRST_CP_COL,   boxplot={hide outliers,draw position=13.10}] table[y=firstcp] {solvecompare_cums20/t500m10r3.dat};  
	\addplot[\LAST_CP_COL,  boxplot={hide outliers,draw position=13.25}] table[y=lastcp]  {solvecompare_cums20/t500m10r3.dat};  

	\addplot[\FIRST_CP_COL,   boxplot={hide outliers,draw position=13.45}] table[y=firstcp]  {solvecompare_cums20/t500m10r5.dat};  
	\addplot[\LAST_CP_COL,  boxplot={hide outliers,draw position=13.60}] table[y=lastcp]   {solvecompare_cums20/t500m10r5.dat};  

	\addplot[\FIRST_CP_COL,   boxplot={hide outliers,draw position=13.75}] table[y=firstcp] {solvecompare_cums20/t500m10r10.dat};  
	\addplot[\LAST_CP_COL,  boxplot={hide outliers,draw position=13.90}] table[y=lastcp]  {solvecompare_cums20/t500m10r10.dat};  
\end{axis}
\end{tikzpicture}
	}
	{
		\insertEmptyFig
	}
	\caption{The differences in schedule execution times produced by the different
	methods for test suites TS1--TS14, with \naive as the baseline of 100\%. 
	The blue is the difference between the first solution \MSPANF and \naive and 
	the red shows the difference between the final solution \MSPANL and \naive.}
	\label{fig:firstsolutionoverview}
\end{figure}

\noindent
To compare our \TSNIC method with the \naive and \rnd methods, we recorded the
first solution, \MSPANF, found by \TSNIC. We also recorded the last solution,
\MSPANL. This is either a proved optimal solution, or the best solution found
after 5~minutes of solving time. For each of the $840$ test suites, we computed
the differences between the \rnd and \naive, \MSPANF and \naive, and \MSPANL and
\naive, where \naive is the baseline of 100\,\%. The results show that \rnd is
30\,\%-60\,\% worse than \naive, which means that \rnd can clearly be discarded
from further analysis. Our findings are summarized in
\figref{fig:firstsolutionoverview}, showing the difference between \TSNIC and
\naive. For all test suites but the hardest subset of TS1 and some instances of
TS2, \MSPANF is better than \naive. We also observe that for larger test suites,
i.e., TS11-TS14, there is only a marginal difference between \MSPANF and
\MSPANL. Hence, running the solver for a longer time has only little benefit.

Furthermore, to evaluate the effectiveness of the test case duration splitting
search strategy, we compared it to standard strategies available in SICStus
Prolog's \clpfd with the same constraint model on the test suites TS1 and TS14.
The search first enumerates on the machine assignments increasingly, i.e.
without load-balancing, and afterwards assigns end times via domain splitting by
bisecting the domain, starting from the earliest end times. As variable
selection strategies, we tested both the default setting, selecting the leftmost
variable, and a first-fail strategy, selecting the variable with the smallest
domain. Additionally, we tried sorting the variables by decreasing resource
usage.

All variants of the standard searches performed substantially worse than test
case duration splitting, with first-fail search on sorted variables being the
best. After finding an initial solution, further improvements are rare and the
makespan of the final solution is in average $4$ times larger compared to using
test case duration splitting with the same time contract of $5$ minutes.

\subsection{RQ2: Will longer solving time reduce the total execution time?}
\label{sec:eval_rq2}
\noindent
RQ2 aims at finding an appropriate trade-off between the time spent in solving
the constraint model, \SOLVINGTIME, and the time spent in executing the
schedule, \MAKESPAN. As mentioned in \secref{sec:intro}, the round-trip time is
critical in \CI and has to be kept low. It is therefore crucial to determine the
most appropriate timeout for the constraint optimizer.
The ultimate goal being to generate a schedule which is quasi-optimal w.r.t.
total execution time, $\TOTALTIME = \SOLVINGTIME + C^*$.

As mentioned above, \TSNIC can be given a time-contract for finding a
quasi-optimal solution when minimizing the execution time of the schedule. More
precisely, with this time-constrained process four outcomes are possible.

\noindent\textbf{No solution with proof}: \TSNIC proves that the OTS problem has no solution due to unsatisfiable constraints.\\
\textbf{No solution without proof}: \TSNIC was not able to find a solution
within the given time. Thus, there could be a solution, but it has not been found.\\
\textbf{Quasi-optimal solution}: At the end of the time-contract, a solution is
returned, but \TSNIC was interrupted while trying to prove its optimality.
Such a best-effort solution is usually sufficient in the examined industrial settings.\\
\textbf{Optimal solution}: Before the end of the time-contract, \TSNIC returns
an optimal solution along with its proof. This is obviously the most desired
result.

Each solution $i$ generated by \TSNIC can be represented by a tuple 
$(\MAKESPAN_{i}, T_{s,i})$ where $\MAKESPAN_{i}$ is the makespan of solution $i$
and $T_{s,i}$ is the time the solver spent finding solution $i$. The goal of RQ2
is to find the value of $T_{s,i}$ that minimizes $(\MAKESPAN_{i} + T_{s,i}), \forall\,i$ 
and use this value as the time-contract.

To answer RQ2, we executed \TSNIC on all $840$ test suites, with a time-contract
of $5$~minutes. During this process, we recorded all intermediate search results
to calculate the optimal value of \SOLVINGTIME for each test suite.

\begin{figure}[ht]
		\ifthenelse{\equal{\includeBigFigures}{true}}
		{
\begin{tikzpicture}
	\def\RND_COL{black}
	\def\FIRST_CP_COL{blue}
	\def\LAST_CP_COL{red}
	\def\RSPACE{\,}
	
\begin{axis}[
  boxplot/draw direction=y,
  boxplot/variable width,
  boxplot/variable width min target=0.07,
  boxplot/every box/.style={fill=gray!50,style=solid},
  boxplot/every whisker/.style={style=solid},
  width=1.0\textwidth,
  height=0.32\textwidth,
  ymode=log,
  xmin= 0,
  xmax=7,
  ymin=1,
  ymax=1000,
  ytick={100,1000},
  yticklabels={0.1,\phantom{50}1},
  xmajorgrids,
  xminorgrids,
  minor tick num=2,
  major grid style={smooth},
  minor grid style={dotted},
  ylabel={Solving time \SOLVINGTIME [s]},
  ylabel near ticks,
  xtick={0,1,2,3,4,5,6,7},
  xticklabels={ { \scriptsize TS1 \\ \scriptsize R3 \RSPACE R5 \RSPACE R10 },    %
  							{ \scriptsize TS2 \\ \scriptsize R3 \RSPACE R5 \RSPACE R10 },    %
  							{ \scriptsize TS3 \\ \scriptsize R3 \RSPACE R5 \RSPACE R10 },    %
  							{ \scriptsize TS4 \\ \scriptsize R3 \RSPACE R5 \RSPACE R10 },    %
  							{ \scriptsize TS5 \\ \scriptsize R3 \RSPACE R5 \RSPACE R10 },    %
  							{ \scriptsize TS6 \\ \scriptsize R3 \RSPACE R5 \RSPACE R10 },    %
  							{ \scriptsize TS7 \\ \scriptsize R3 \RSPACE R5 \RSPACE R10 }},   %
  x tick label style={ yshift = -1.2em  },
  x tick label as interval,
  xticklabel pos=upper,
  xticklabel style={align=center}
  ]
  
	\addplot[mark=none, black] coordinates { (0,1) (8.5,1) };

	\addplot+[\RND_COL,   boxplot={hide outliers, draw position=0.05}] table[y=tfirst]  {firstlast_cums20/t20m10r3_f_l.dat};  
	\addplot+[\FIRST_CP_COL, boxplot={hide outliers, draw position=0.15}] table[y=optforTt] {firstlast_cums20/t20m10r3_f_l.dat};  
	\addplot+[\LAST_CP_COL,  boxplot={hide outliers, draw position=0.25}] table[y=tlast]  {firstlast_cums20/t20m10r3_f_l.dat};  
	
	\addplot+[\RND_COL,     boxplot={hide outliers, draw position=0.4}] table[y=tfirst]   {firstlast_cums20/t20m10r5_f_l.dat};  
	\addplot+[\FIRST_CP_COL, boxplot={hide outliers, draw position=0.5}] table[y=optforTt]  {firstlast_cums20/t20m10r5_f_l.dat};  
	\addplot+[\LAST_CP_COL,  boxplot={hide outliers, draw position=0.6}] table[y=tlast]   {firstlast_cums20/t20m10r5_f_l.dat};  
	\addplot+[\RND_COL,    boxplot={hide outliers, draw position=0.75}] table[y=tfirst]  {firstlast_cums20/t20m10r10_f_l.dat};  
	\addplot+[\FIRST_CP_COL,   boxplot={hide outliers, draw position=0.85}] table[y=optforTt] {firstlast_cums20/t20m10r10_f_l.dat};  
	\addplot+[\LAST_CP_COL,  boxplot={hide outliers, draw position=0.95}] table[y=tlast]  {firstlast_cums20/t20m10r10_f_l.dat};  
	\addplot+[\RND_COL,    boxplot={hide outliers, draw position=1.05}] table[y=tfirst]  {firstlast_cums20/t30m20r3_f_l.dat};  
	\addplot+[\FIRST_CP_COL,   boxplot={hide outliers, draw position=1.15}] table[y=optforTt] {firstlast_cums20/t30m20r3_f_l.dat};  
	\addplot+[\LAST_CP_COL,  boxplot={hide outliers, draw position=1.25}] table[y=tlast]  {firstlast_cums20/t30m20r3_f_l.dat};  
	\addplot+[\RND_COL,    boxplot={hide outliers, draw position=1.4}] table[y=tfirst]   {firstlast_cums20/t30m20r5_f_l.dat};  
	\addplot+[\FIRST_CP_COL,   boxplot={hide outliers, draw position=1.5}] table[y=optforTt]  {firstlast_cums20/t30m20r5_f_l.dat};  
	\addplot+[\LAST_CP_COL,  boxplot={hide outliers, draw position=1.6}] table[y=tlast]   {firstlast_cums20/t30m20r5_f_l.dat};  
	\addplot+[\RND_COL,    boxplot={hide outliers, draw position=1.75}] table[y=tfirst]  {firstlast_cums20/t30m20r10_f_l.dat};  
	\addplot+[\FIRST_CP_COL,   boxplot={hide outliers, draw position=1.85}] table[y=optforTt] {firstlast_cums20/t30m20r10_f_l.dat};  
	\addplot+[\LAST_CP_COL,  boxplot={hide outliers, draw position=1.95}] table[y=tlast]  {firstlast_cums20/t30m20r10_f_l.dat};  
	\addplot+[\RND_COL,    boxplot={hide outliers, draw position=2.05}] table[y=tfirst]  {firstlast_cums20/t30m10r3_f_l.dat};  
	\addplot+[\FIRST_CP_COL,   boxplot={hide outliers, draw position=2.15}] table[y=optforTt] {firstlast_cums20/t30m10r3_f_l.dat};  
	\addplot+[\LAST_CP_COL,  boxplot={hide outliers, draw position=2.25}] table[y=tlast]  {firstlast_cums20/t30m10r3_f_l.dat};  
	\addplot+[\RND_COL,    boxplot={hide outliers, draw position=2.4}] table[y=tfirst]   {firstlast_cums20/t30m10r5_f_l.dat};  
	\addplot+[\FIRST_CP_COL,   boxplot={hide outliers, draw position=2.5}] table[y=optforTt]  {firstlast_cums20/t30m10r5_f_l.dat};  
	\addplot+[\LAST_CP_COL,  boxplot={hide outliers, draw position=2.6}] table[y=tlast]   {firstlast_cums20/t30m10r5_f_l.dat};  
	\addplot+[\RND_COL,    boxplot={hide outliers, draw position=2.75}] table[y=tfirst]  {firstlast_cums20/t30m10r10_f_l.dat};  
	\addplot+[\FIRST_CP_COL,   boxplot={hide outliers, draw position=2.85}] table[y=optforTt] {firstlast_cums20/t30m10r10_f_l.dat};  
	\addplot+[\LAST_CP_COL,  boxplot={hide outliers, draw position=2.95}] table[y=tlast]  {firstlast_cums20/t30m10r10_f_l.dat};  
	\addplot+[\RND_COL,    boxplot={hide outliers, draw position=3.05}] table[y=tfirst]  {firstlast_cums20/t40m20r3_f_l.dat};  
	\addplot+[\FIRST_CP_COL,   boxplot={hide outliers, draw position=3.15}] table[y=optforTt] {firstlast_cums20/t40m20r3_f_l.dat};  
	\addplot+[\LAST_CP_COL,  boxplot={hide outliers, draw position=3.25}] table[y=tlast]  {firstlast_cums20/t40m20r3_f_l.dat};  
	\addplot+[\RND_COL,    boxplot={hide outliers, draw position=3.4}] table[y=tfirst]   {firstlast_cums20/t40m20r5_f_l.dat};  
	\addplot+[\FIRST_CP_COL,   boxplot={hide outliers, draw position=3.5}] table[y=optforTt]  {firstlast_cums20/t40m20r5_f_l.dat};  
	\addplot+[\LAST_CP_COL,  boxplot={hide outliers, draw position=3.6}] table[y=tlast]   {firstlast_cums20/t40m20r5_f_l.dat};  
	\addplot+[\RND_COL,    boxplot={hide outliers, draw position=3.75}] table[y=tfirst]  {firstlast_cums20/t40m20r10_f_l.dat};  
	\addplot+[\FIRST_CP_COL,   boxplot={hide outliers, draw position=3.85}] table[y=optforTt] {firstlast_cums20/t40m20r10_f_l.dat};  
	\addplot+[\LAST_CP_COL,  boxplot={hide outliers, draw position=3.95}] table[y=tlast]  {firstlast_cums20/t40m20r10_f_l.dat};  
	\addplot+[\RND_COL,    boxplot={hide outliers, draw position=4.05}] table[y=tfirst]  {firstlast_cums20/t40m10r3_f_l.dat};  
	\addplot+[\FIRST_CP_COL,   boxplot={hide outliers, draw position=4.15}] table[y=optforTt] {firstlast_cums20/t40m10r3_f_l.dat};  
	\addplot+[\LAST_CP_COL,  boxplot={hide outliers, draw position=4.25}] table[y=tlast]  {firstlast_cums20/t40m10r3_f_l.dat};  
	\addplot+[\RND_COL,    boxplot={hide outliers, draw position=4.4}] table[y=tfirst]   {firstlast_cums20/t40m10r5_f_l.dat};  
	\addplot+[\FIRST_CP_COL,   boxplot={hide outliers, draw position=4.5}] table[y=optforTt]  {firstlast_cums20/t40m10r5_f_l.dat};  
	\addplot+[\LAST_CP_COL,  boxplot={hide outliers, draw position=4.6}] table[y=tlast]   {firstlast_cums20/t40m10r5_f_l.dat};  
	\addplot+[\RND_COL,    boxplot={hide outliers, draw position=4.75}] table[y=tfirst]  {firstlast_cums20/t40m10r10_f_l.dat};  
	\addplot+[\FIRST_CP_COL,   boxplot={hide outliers, draw position=4.85}] table[y=optforTt] {firstlast_cums20/t40m10r10_f_l.dat};  
	\addplot+[\LAST_CP_COL,  boxplot={hide outliers, draw position=4.95}] table[y=tlast]  {firstlast_cums20/t40m10r10_f_l.dat};  
	\addplot+[\RND_COL,    boxplot={hide outliers, draw position=5.05}] table[y=tfirst]  {firstlast_cums20/t50m20r3_f_l.dat};  
	\addplot+[\FIRST_CP_COL,   boxplot={hide outliers, draw position=5.15}] table[y=optforTt] {firstlast_cums20/t50m20r3_f_l.dat};  
	\addplot+[\LAST_CP_COL,  boxplot={hide outliers, draw position=5.25}] table[y=tlast]  {firstlast_cums20/t50m20r3_f_l.dat};  
	\addplot+[\RND_COL,    boxplot={hide outliers, draw position=5.4}] table[y=tfirst]   {firstlast_cums20/t50m20r5_f_l.dat};  
	\addplot+[\FIRST_CP_COL,   boxplot={hide outliers, draw position=5.5}] table[y=optforTt]  {firstlast_cums20/t50m20r5_f_l.dat};  
	\addplot+[\LAST_CP_COL,  boxplot={hide outliers, draw position=5.6}] table[y=tlast]   {firstlast_cums20/t50m20r5_f_l.dat};  
	\addplot+[\RND_COL,    boxplot={hide outliers, draw position=5.75}] table[y=tfirst]  {firstlast_cums20/t50m20r10_f_l.dat};  
	\addplot+[\FIRST_CP_COL,   boxplot={hide outliers, draw position=5.85}] table[y=optforTt] {firstlast_cums20/t50m20r10_f_l.dat};  
	\addplot+[\LAST_CP_COL,  boxplot={hide outliers, draw position=5.95}] table[y=tlast]  {firstlast_cums20/t50m20r10_f_l.dat};  
	\addplot+[\RND_COL,    boxplot={hide outliers, draw position=6.05}] table[y=tfirst]  {firstlast_cums20/t50m10r3_f_l.dat};  
	\addplot+[\FIRST_CP_COL,   boxplot={hide outliers, draw position=6.15}] table[y=optforTt] {firstlast_cums20/t50m10r3_f_l.dat};  
	\addplot+[\LAST_CP_COL,  boxplot={hide outliers, draw position=6.25}] table[y=tlast]  {firstlast_cums20/t50m10r3_f_l.dat};  
	\addplot+[\RND_COL,    boxplot={hide outliers, draw position=6.4}] table[y=tfirst]   {firstlast_cums20/t50m10r5_f_l.dat};  
	\addplot+[\FIRST_CP_COL,   boxplot={hide outliers, draw position=6.5}] table[y=optforTt]  {firstlast_cums20/t50m10r5_f_l.dat};  
	\addplot+[\LAST_CP_COL,  boxplot={hide outliers, draw position=6.6}] table[y=tlast]   {firstlast_cums20/t50m10r5_f_l.dat};  
	\addplot+[\RND_COL,    boxplot={hide outliers, draw position=6.75}] table[y=tfirst]  {firstlast_cums20/t50m10r10_f_l.dat};  
	\addplot+[\FIRST_CP_COL,   boxplot={hide outliers, draw position=6.85}] table[y=optforTt] {firstlast_cums20/t50m10r10_f_l.dat};  
	\addplot+[\LAST_CP_COL,  boxplot={hide outliers, draw position=6.95}] table[y=tlast]  {firstlast_cums20/t50m10r10_f_l.dat};  

\end{axis}

\end{tikzpicture} 			\vspace{-0.8em}

\begin{tikzpicture}
	\def\RND_COL{black}
	\def\FIRST_CP_COL{blue}
	\def\LAST_CP_COL{red}
	\def\RSPACE{\,}
		
\begin{axis}[
  boxplot/draw direction=y,
  boxplot/variable width,
  boxplot/variable width min target=0.07,
  boxplot/every box/.style={fill=gray!50},
  width=1.0\textwidth,
  height=0.32\textwidth,
  ymode=log,
  xmin= 7,
  xmax=14,
  ymin=100,
  ymax=500000,
  ytick={1000, 10000, 60000, 500000},
  yticklabels={1, 10, 60, 500},
  xmajorgrids,
  xminorgrids,
  minor tick num=2,
  major grid style={smooth},
  minor grid style={dotted},
  ylabel={Solving time \SOLVINGTIME [s]},
  ylabel near ticks,
  xtick={7,8,9,10,11,12,13,14},
  xticklabels={ { \scriptsize TS8  \\ \scriptsize R3 \RSPACE R5 \RSPACE R10 },    %
  							{ \scriptsize TS9  \\ \scriptsize R3 \RSPACE R5 \RSPACE R10 },    %
  							{ \scriptsize TS10 \\ \scriptsize R3 \RSPACE R5 \RSPACE R10 },    %
  							{ \scriptsize TS11 \\ \scriptsize R3 \RSPACE R5 \RSPACE R10 },    %
  							{ \scriptsize TS12 \\ \scriptsize R3 \RSPACE R5 \RSPACE R10 },    %
  							{ \scriptsize TS13 \\ \scriptsize R3 \RSPACE R5 \RSPACE R10 },    %
  							{ \scriptsize TS14 \\ \scriptsize R3 \RSPACE R5 \RSPACE R10 }},   %
  x tick label style={ yshift = -1.2em  },
  x tick label as interval,
  xticklabel pos=upper,
  xticklabel style={align=center}
  ]
                  
	\addplot[mark=none, black] coordinates { (0,1) (21.5,1) };

	\addplot[\RND_COL,    boxplot={hide outliers,draw position=7.05}] table[y=tfirst]  {firstlast_cums20/t100m50r3_f_l.dat};  
	\addplot[\FIRST_CP_COL,   boxplot={hide outliers,draw position=7.15}] table[y=optforTt] {firstlast_cums20/t100m50r3_f_l.dat};  
	\addplot[\LAST_CP_COL,  boxplot={hide outliers,draw position=7.25}] table[y=tlast]  {firstlast_cums20/t100m50r3_f_l.dat};  
	\addplot[\RND_COL,    boxplot={hide outliers,draw position=7.4}] table[y=tfirst]   {firstlast_cums20/t100m50r5_f_l.dat};  
	\addplot[\FIRST_CP_COL,   boxplot={hide outliers,draw position=7.5}] table[y=optforTt]  {firstlast_cums20/t100m50r5_f_l.dat};  
	\addplot[\LAST_CP_COL,  boxplot={hide outliers,draw position=7.6}] table[y=tlast]   {firstlast_cums20/t100m50r5_f_l.dat};  
	\addplot[\RND_COL,    boxplot={hide outliers,draw position=7.75}] table[y=tfirst]  {firstlast_cums20/t100m50r10_f_l.dat};  
	\addplot[\FIRST_CP_COL,   boxplot={hide outliers,draw position=7.85}] table[y=optforTt] {firstlast_cums20/t100m50r10_f_l.dat};  
	\addplot[\LAST_CP_COL,  boxplot={hide outliers,draw position=7.95}] table[y=tlast]  {firstlast_cums20/t100m50r10_f_l.dat};

	\addplot[\RND_COL,    boxplot={hide outliers,draw position=8.05}] table[y=tfirst]  {firstlast_cums20/t100m20r3_f_l.dat};  
	\addplot[\FIRST_CP_COL,   boxplot={hide outliers,draw position=8.15}] table[y=optforTt] {firstlast_cums20/t100m20r3_f_l.dat};  
	\addplot[\LAST_CP_COL,  boxplot={hide outliers,draw position=8.25}] table[y=tlast]  {firstlast_cums20/t100m20r3_f_l.dat};  
	\addplot[\RND_COL,    boxplot={hide outliers,draw position=8.4}] table[y=tfirst]   {firstlast_cums20/t100m20r5_f_l.dat};  
	\addplot[\FIRST_CP_COL,   boxplot={hide outliers,draw position=8.5}] table[y=optforTt]  {firstlast_cums20/t100m20r5_f_l.dat};  
	\addplot[\LAST_CP_COL,  boxplot={hide outliers,draw position=8.6}] table[y=tlast]   {firstlast_cums20/t100m20r5_f_l.dat};  
	\addplot[\RND_COL,    boxplot={hide outliers,draw position=8.75}] table[y=tfirst]  {firstlast_cums20/t100m20r10_f_l.dat};  
	\addplot[\FIRST_CP_COL,   boxplot={hide outliers,draw position=8.85}] table[y=optforTt] {firstlast_cums20/t100m20r10_f_l.dat};  
	\addplot[\LAST_CP_COL,  boxplot={hide outliers,draw position=8.95}] table[y=tlast]  {firstlast_cums20/t100m20r10_f_l.dat};  
	\addplot[\RND_COL,    boxplot={hide outliers,draw position=9.05}] table[y=tfirst]  {firstlast_cums20/t100m10r3_f_l.dat};  
	\addplot[\FIRST_CP_COL,   boxplot={hide outliers,draw position=9.15}] table[y=optforTt] {firstlast_cums20/t100m10r3_f_l.dat};  
	\addplot[\LAST_CP_COL,  boxplot={hide outliers,draw position=9.25}] table[y=tlast]  {firstlast_cums20/t100m10r3_f_l.dat};  
	\addplot[\RND_COL,    boxplot={hide outliers,draw position=9.4}] table[y=tfirst]   {firstlast_cums20/t100m10r5_f_l.dat};  
	\addplot[\FIRST_CP_COL,   boxplot={hide outliers,draw position=9.5}] table[y=optforTt]  {firstlast_cums20/t100m10r5_f_l.dat};  
	\addplot[\LAST_CP_COL,  boxplot={hide outliers,draw position=9.6}] table[y=tlast]   {firstlast_cums20/t100m10r5_f_l.dat};  
	\addplot[\RND_COL,    boxplot={hide outliers,draw position=9.75}] table[y=tfirst]  {firstlast_cums20/t100m10r10_f_l.dat};  
	\addplot[\FIRST_CP_COL,   boxplot={hide outliers,draw position=9.85}] table[y=optforTt] {firstlast_cums20/t100m10r10_f_l.dat};  
	\addplot[\LAST_CP_COL,  boxplot={hide outliers,draw position=9.95}] table[y=tlast]  {firstlast_cums20/t100m10r10_f_l.dat};  
	\addplot[\RND_COL,    boxplot={hide outliers,draw position=10.05}] table[y=tfirst]  {firstlast_cums20/t500m100r3_f_l.dat};  
	\addplot[\FIRST_CP_COL,   boxplot={hide outliers,draw position=10.15}] table[y=optforTt] {firstlast_cums20/t500m100r3_f_l.dat};  
	\addplot[\LAST_CP_COL,  boxplot={hide outliers,draw position=10.25}] table[y=tlast]  {firstlast_cums20/t500m100r3_f_l.dat};  
	\addplot[\RND_COL,    boxplot={hide outliers,draw position=10.4}] table[y=tfirst]   {firstlast_cums20/t500m100r5_f_l.dat};  
	\addplot[\FIRST_CP_COL,   boxplot={hide outliers,draw position=10.5}] table[y=optforTt]  {firstlast_cums20/t500m100r5_f_l.dat};  
	\addplot[\LAST_CP_COL,  boxplot={hide outliers,draw position=10.6}] table[y=tlast]   {firstlast_cums20/t500m100r5_f_l.dat};  
	\addplot[\RND_COL,    boxplot={hide outliers,draw position=10.75}] table[y=tfirst]  {firstlast_cums20/t500m100r10_f_l.dat};  
	\addplot[\FIRST_CP_COL,   boxplot={hide outliers,draw position=10.85}] table[y=optforTt] {firstlast_cums20/t500m100r10_f_l.dat};  
	\addplot[\LAST_CP_COL,  boxplot={hide outliers,draw position=10.95}] table[y=tlast]  {firstlast_cums20/t500m100r10_f_l.dat};  
	\addplot[\RND_COL,    boxplot={hide outliers,draw position=11.05}] table[y=tfirst]  {firstlast_cums20/t500m50r3_f_l.dat};  
	\addplot[\FIRST_CP_COL,   boxplot={hide outliers,draw position=11.15}] table[y=optforTt] {firstlast_cums20/t500m50r3_f_l.dat};  
	\addplot[\LAST_CP_COL,  boxplot={hide outliers,draw position=11.25}] table[y=tlast]  {firstlast_cums20/t500m50r3_f_l.dat};  
	\addplot[\RND_COL,    boxplot={hide outliers,draw position=11.4}] table[y=tfirst]   {firstlast_cums20/t500m50r5_f_l.dat};  
	\addplot[\FIRST_CP_COL,   boxplot={hide outliers,draw position=11.5}] table[y=optforTt]  {firstlast_cums20/t500m50r5_f_l.dat};  
	\addplot[\LAST_CP_COL,  boxplot={hide outliers,draw position=11.6}] table[y=tlast]   {firstlast_cums20/t500m50r5_f_l.dat};  
	\addplot[\RND_COL,    boxplot={hide outliers,draw position=11.75}] table[y=tfirst]  {firstlast_cums20/t500m50r10_f_l.dat};  
	\addplot[\FIRST_CP_COL,   boxplot={hide outliers,draw position=11.85}] table[y=optforTt] {firstlast_cums20/t500m50r10_f_l.dat};  
	\addplot[\LAST_CP_COL,  boxplot={hide outliers,draw position=11.95}] table[y=tlast]  {firstlast_cums20/t500m50r10_f_l.dat};  
	\addplot[\RND_COL,    boxplot={hide outliers,draw position=12.05}] table[y=tfirst]  {firstlast_cums20/t500m20r3_f_l.dat};  
	\addplot[\FIRST_CP_COL,   boxplot={hide outliers,draw position=12.15}] table[y=optforTt] {firstlast_cums20/t500m20r3_f_l.dat};  
	\addplot[\LAST_CP_COL,  boxplot={hide outliers,draw position=12.25}] table[y=tlast]  {firstlast_cums20/t500m20r3_f_l.dat};  
	\addplot[\RND_COL,    boxplot={hide outliers,draw position=12.4}] table[y=tfirst]   {firstlast_cums20/t500m20r5_f_l.dat};  
	\addplot[\FIRST_CP_COL,   boxplot={hide outliers,draw position=12.5}] table[y=optforTt]  {firstlast_cums20/t500m20r5_f_l.dat};  
	\addplot[\LAST_CP_COL,  boxplot={hide outliers,draw position=12.6}] table[y=tlast]   {firstlast_cums20/t500m20r5_f_l.dat};  
	\addplot[\RND_COL,    boxplot={hide outliers,draw position=12.75}] table[y=tfirst]  {firstlast_cums20/t500m20r10_f_l.dat};  
	\addplot[\FIRST_CP_COL,   boxplot={hide outliers,draw position=12.85}] table[y=optforTt] {firstlast_cums20/t500m20r10_f_l.dat};  
	\addplot[\LAST_CP_COL,  boxplot={hide outliers,draw position=12.95}] table[y=tlast]  {firstlast_cums20/t500m20r10_f_l.dat};  
	\addplot[\RND_COL,    boxplot={hide outliers,draw position=13.05}] table[y=tfirst]  {firstlast_cums20/t500m10r3_f_l.dat};  
	\addplot[\FIRST_CP_COL,   boxplot={hide outliers,draw position=13.15}] table[y=optforTt] {firstlast_cums20/t500m10r3_f_l.dat};  
	\addplot[\LAST_CP_COL,  boxplot={hide outliers,draw position=13.25}] table[y=tlast]  {firstlast_cums20/t500m10r3_f_l.dat};  
	\addplot[\RND_COL,    boxplot={hide outliers,draw position=13.4}] table[y=tfirst]   {firstlast_cums20/t500m10r5_f_l.dat};  
	\addplot[\FIRST_CP_COL,   boxplot={hide outliers,draw position=13.5}] table[y=optforTt]  {firstlast_cums20/t500m10r5_f_l.dat};  
	\addplot[\LAST_CP_COL,  boxplot={hide outliers,draw position=13.6}] table[y=tlast]   {firstlast_cums20/t500m10r5_f_l.dat};  
	\addplot[\RND_COL,    boxplot={hide outliers,draw position=13.75}] table[y=tfirst]  {firstlast_cums20/t500m10r10_f_l.dat};  
	\addplot[\FIRST_CP_COL,   boxplot={hide outliers,draw position=13.85}] table[y=optforTt] {firstlast_cums20/t500m10r10_f_l.dat};  
	\addplot[\LAST_CP_COL,  boxplot={hide outliers,draw position=13.95}] table[y=tlast]  {firstlast_cums20/t500m10r10_f_l.dat};  
\end{axis}
\end{tikzpicture} 			%
		}
		{		
			\insertEmptyFig
		}
		
		\caption{The black boxes show the distribution in solving time, \SOLVINGTIME, for the first 
		solution found by \TSNIC. The blue boxes show the distribution in \SOLVINGTIME where the total 
		execution time, \TOTALTIME, is optimal. Finally, the red boxes show the distribution in
		\SOLVINGTIME for the last solution found by \TSNIC, which can be the optimal value or the last
		value found before timeout. The timeout was set to 5\,min.}
		\label{fig:firstoptlast}
		\CAPSQUIZE
	\end{figure}
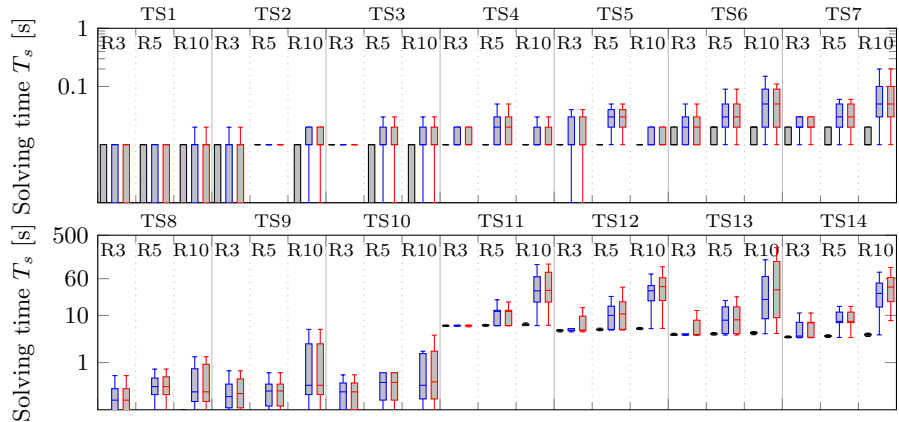

\figref{fig:firstoptlast} shows the
distribution in solving time for the first solution found by \TSNIC, the last
solution and also how the optimal value of \SOLVINGTIME is distributed. For the
group of $600$ test suites containing up to $100$ test cases (TS1-TS10), the
results show that a solution that minimizes the total execution time, noted
\TOTALTIME, is found in $\SOLVINGTIME < 5$\,s for $96.8\,\%$ of the test suites.
If we extend the search time to $\SOLVINGTIME < 10$\,s, the number grows to
$98\,\%$ of the test suites. For this group, the worst case optimal solving time
was $\SOLVINGTIME=122.3$\,s. We see that a solution is always found in less than
$0.1$\,s. For the group of $240$ test suites containing $500$ test cases
(TS11-TS14), the results show that a solution that minimizes \TOTALTIME is found
in $\SOLVINGTIME < 120$\,s for $97.5\,\%$ of the test suites. A solution
minimizing \TOTALTIME is found in less than 240\,s for all test suites, except
one instance with $\TOTALTIME = 264$\,s.

  An increased investment in the solving part does not seem to necessarily pay
  off if one considers the total execution time. The reported experiments
  give hints to evaluate and select the optimal test contract for the solving
  part.

\subsection{RQ3: Can \TSNIC efficiently solve industrial OTS problems?}
\noindent
To answer RQ3, we consider two of the three industrial case studies, namely, IPS
and VCS. These case studies are composed of automated test scripts, which makes
the application of the \TSNIC method especially pertinent.

In both case studies, the guaranteed optimal solution is already found as the
first solution in less than $200$~ms. This avoids the necessity to compromise
between $\MAKESPAN$ and $\SOLVINGTIME$ for these industrial applications.

When applying \TSNIC to the IPS test suite, we find the optimal solution,
$\MAKESPAN=780$~s, at $\SOLVINGTIME=10$~ms. For the VCS test suite, the optimal
solution, $\MAKESPAN=14637$~s is found at $\SOLVINGTIME=160$~ms.

In summary, \TSNIC can easily be applied to both VCS and IPS, and in both cases,
the best result is achieved when $\MAKESPAN$ is minimized and $\SOLVINGTIME$ is
neglected.

\section{Conclusion}
\label{sec:conclusion}
\noindent
This paper introduced \TSNIC, a time-aware method for solving the optimal test
suite scheduling (OTS) problem, where test cases can be executed on multiple
execution machines with non-shareable global resources. \TSNIC exploits the
\Cumulatives global constraint and a time-aware minimization process, and a
dedicated search strategy, called {\it test case duration splitting}. To our
knowledge, the OTS problem is rigorously formalized for the first time and a
method is proposed to solve it in \CI applications. An experimental evaluation
performed over $840$ generated test suites revealed that \TSNIC outperforms
simple scheduling methods w.r.t. total execution time. More specifically, we
showed that automatic optimal scheduling of $500$ test cases over $100$ machines
is reachable in less than 4 minutes for $99.5\%$ instances of the problem. By
considering trade-offs between the solving time and the total execution time,
the evaluation allowed us to find the best compromise to allocate time-contracts
to the solving process. Finally, by using \TSNIC with two industrial test
suites, we demonstrated that finding the guaranteed optimal test execution time
is possible and that \TSNIC can effectively solve the OTS problem in practice.

Further work includes consideration of test case priorities, non-unitary
shareable global resources, as well as explicit symmetry breaking in the model.
Additional evaluation and comparison against heuristic methods, such as
evolutionary algorithms, or Mixed-Integer Linear Programming could extend the
presented work and support the integration of \TSNIC in practical \CI processes.

\bibliographystyle{splncs03}
\normalem
\bibliography{ms}

\end{document}